\documentclass[11pt,twoside]{article}
\usepackage{asp2010}
\usepackage{epsfig}
\usepackage{graphicx}
\resetcounters

\begin{document}

\title{The Effect of Environment on Massive Star Formation}
\author{Rowan J. Smith$^1$, Paul C. Clark$^1$, Simon C. O. Glover$^1$, Ian A. Bonnell$^2$, and Ralf S. Klessen$^{1,3}$ 
\affil{$^1$ Zentrum f\"ur Astronomie der Universit\"at Heidelberg, Institut f\"ur Theoretische Astrophysik, Albert-Ueberle-Str. 2, 69120 Heidelberg, Germany }
\affil{$^2$ SUPA, School of Physics \& Astronomy, University of St Andrews, North Haugh, St Andrews, Fife, KY16 9SS, UK }
\affil{$^3$ Kavli Institute for Particle Astrophysics and Cosmology, Stanford University, Menlo Park, CA 94025, USA}}

\begin{abstract}
In this contribution we review our recent numerical work discussing the essential role of the local cluster environment in assembling massive stars. First we show that massive stars are formed from low mass pre-stellar cores and become massive due to accretion. Proto-stars that benefit from this accretion are those situated at the centre of a cluster's potential well, which is the focal point of the contraction of the cluster gas. Given that most of the mass which makes up a massive star in this model comes from the cluster environment rather than the core, it is important to model the molecular cloud environment accurately. Preliminary results of a simulation which accurately treats the chemistry and time-dependent thermodynamics of a molecular cloud show quantitatively similar star formation to previous models, but allow a true comparison to be made between simulation and observations. This method can also be applied to cases with varying metallicities allowing star formati
 on in primordial gas to be studied. In general, these numerical studies of clustered star formation yield IMFs which are compatible with the Salpeter mass function. The only possible exception to this is in low density unbound regions of molecular clouds which lack very low and high mass stars.
\end{abstract}

\section{The Assembly of Massive Stars} 

\subsection{Are massive stars formed from dense cores?}
Since the observations of \citet{Motte98}, it has been noted that the mass function of dense cores, closely resembles that of the stellar initial mass function \citep[eg.][]{Johnstone00,Testi98}. This has lead many to propose that there is a direct link between the two distributions, with the mass of the core uniquely determining the mass of the final stellar system formed from it \citep{Alves07,Simpson08}. In this scenario massive stars would be formed from massive pre-stellar cores. However, the Jeans mass in molecular clouds is typically less than a solar mass, so it is unclear how such cores would be supported against further fragmentation. This could perhaps be resolved with turbulent support, as proposed in \citet{McKee03}. However, turbulence creates density enhancements which would themselves rapidly collapse \citep{Dobbs05}. One way of overcoming this obstacle would be to invoke the effects of radiative heating from surrounding small protostars, which would raise the
  Jeans mass locally \citep{Krumholz08}. However this model is analytic and does not explicitly model the cluster environment. It does not explain why the region of gas surrounded by the low mass stars had previously avoided collapse, despite being centrally condensed and at high densities.

The connection between the core mass function and the IMF was probed in \citet{Smith09} using a hydrodynamic SPH simulation (without feedback) of a giant molecular cloud. In this paper, gravitationally bound `p-cores' were identified using a clump-finding algorithm that finds structures which are part of a common gravitational potential minimum. `P-cores' represent the bound cores which will unambiguously proceed to form stars, and as such are a best-case scenario when compared to observational cores. The cores identified in observations of real clouds suffer from a variety of effects, such as confusion and mass uncertainties \citep{Kainulainen09,Pineda08,diFrancesco07} that make the determination of the masses of bound objects far more difficult. Since in this simulation the final outcome of the star-formation process is known, i.e. the mass that is converted sink particles \citep{Bate95}, it is possible to directly trace the link between the two populations.

\begin{figure}
\plottwo{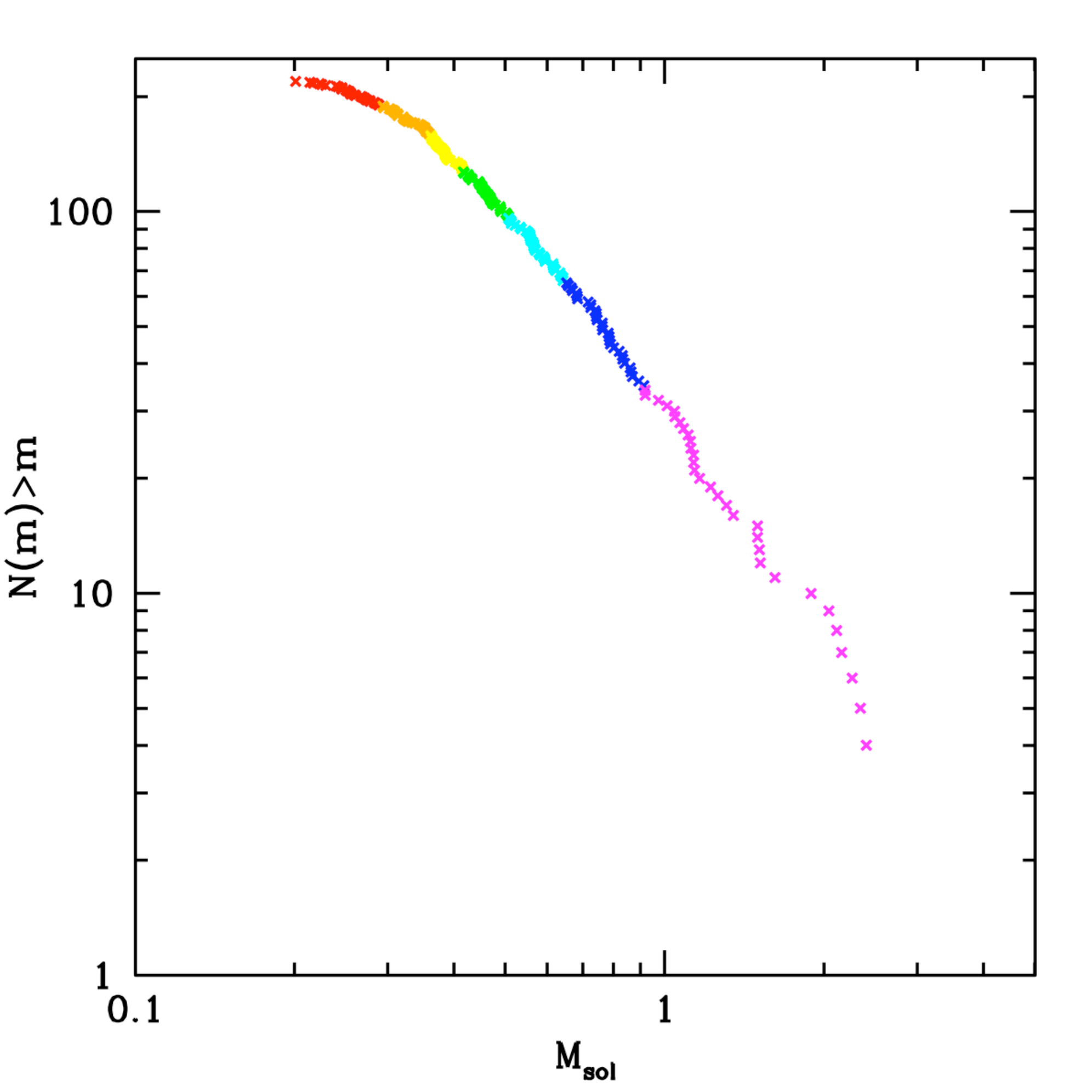}{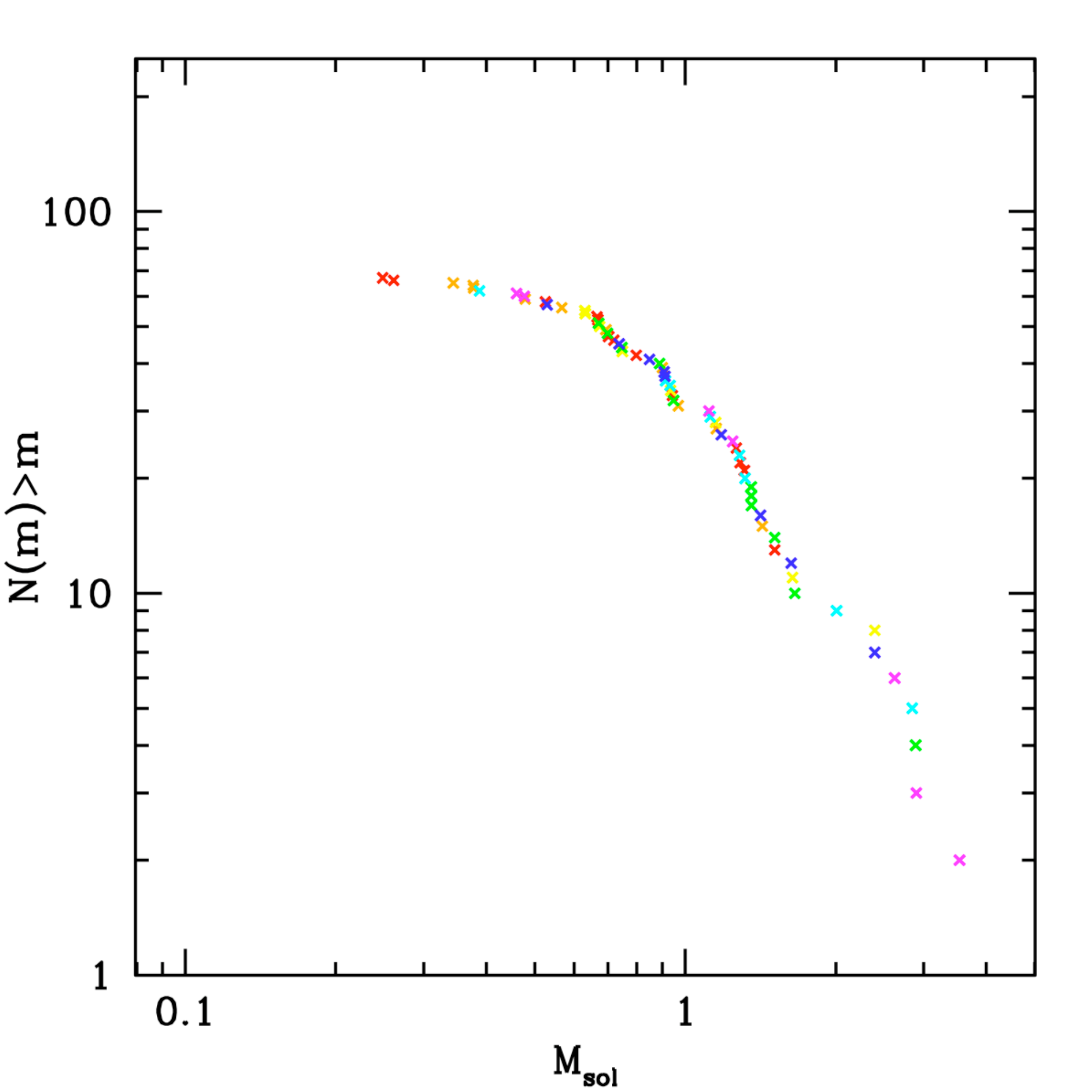}
\caption{\textit{Left} The cumulative mass function of p-cores, in which the mass of the p-core is recorded at the time it is first bound. Each point represents a single core and is colour coded by mass. \textit{Right} The cumulative mass function of the resulting sink particles formed from the cores after they have evolved for five free-fall times. Each point on the graph represents the total mass of sinks formed from the cores and the colour coding is kept the same as in the previous plot. If the stellar mass function were determined solely by the core mass function, one would expect that points of the same colour would still be grouped together. However, this is not the case.
\label{fig:core_mf}
}
\end{figure}

The left panel of Figure \ref{fig:core_mf} shows the initial cumulative mass function of p-cores at the point that they first become gravitationally bound. The points represent individual cores and are colour coded by mass. The right panel shows the mass of the stellar systems formed from these cores after five free-fall times with the same colour coding. If the stellar mass function were determined solely by the core mass function, one would expect to find that points of the same colour would still be grouped together. However, this is not the case. The points are disordered and accretion from the local environment has allowed many stellar systems to grow larger than their original precursor core mass.
Even more significantly, in this simulation there are no massive cores, yet at later times in the simulation massive stars are formed. There are three stars in this simulation with masses over $15$ M$_{\odot}$; so if not from cores, then how is the mass assembled?

\subsection{Where does the mass come from?}
\begin{figure}
\plottwo{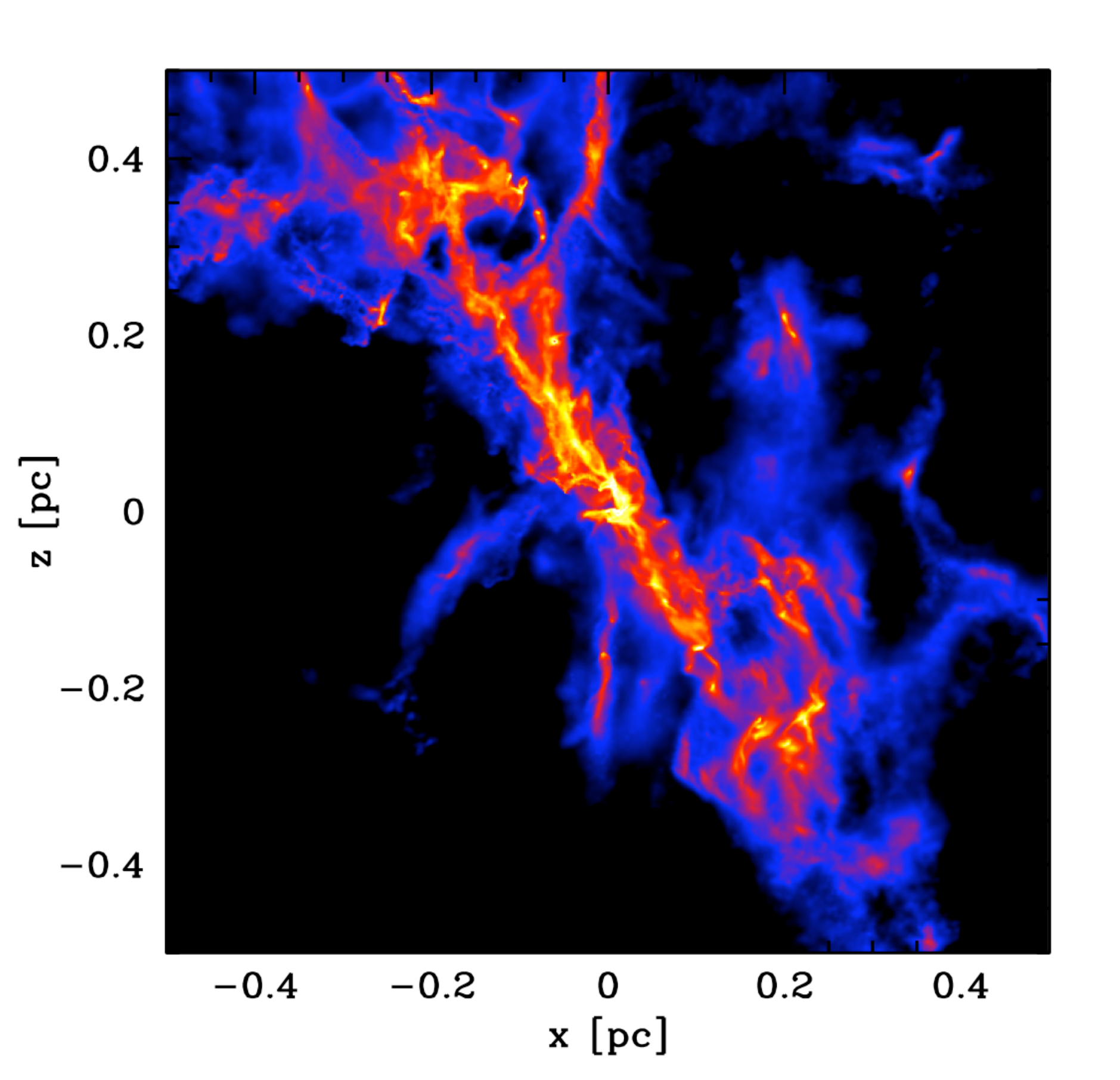}{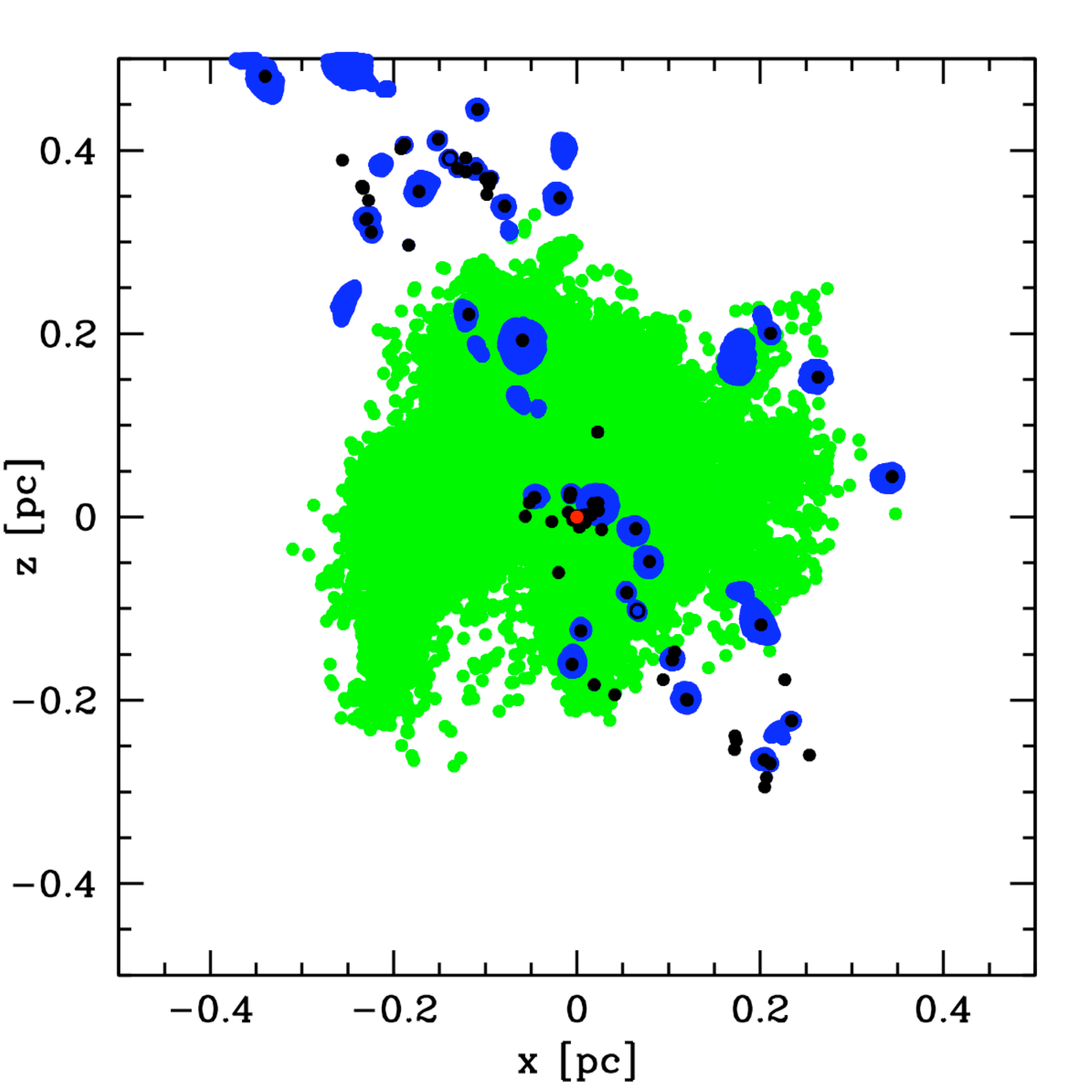}
\caption{\textit{Left} The column density of gas in a cluster forming clump, at the centre of which a massive star is forming. The colour scale runs from $0.05$ g cm$^{-2}$ to $5$ g cm$^{-2}$. 
\textit{Right} The final fate of the gas. The green dots show the positions of gas which will eventually be accreted by the massive sink (red dot). Black dots show the position of sinks and blue dots show the location of material in cores. The gas which will be accreted by the massive sinks is well distributed throughout the clump, and generally cores within this region will not be disrupted by the massive sink.
\label{fig:fate}
}
\end{figure}

This issue was addressed in \citet{Smith09b}. In a higher resolution version of the previous simulation which included an approximation of thermal feedback from the sinks the evolution of the clumps of gas surrounding three massive sink particles was traced with time. The left hand panel of Figure \ref{fig:fate} shows the column density of a large clump of gas which is forming a large stellar cluster. At the centre of this region a massive star is forming which has a mass of $29.2$ M$_{\odot}$ at the end of the simulation. The right hand panel of Figure \ref{fig:fate} shows the position of gas which will be accreted by the massive star in the next 100,000 yr. 
The mass that goes into the massive star does not come from a well-defined `core', but instead is distributed throughout the larger clump volume. Within this volume there also exist dense cores of gas (shown in blue) but the material accreted by the most massive star is the more diffuse inter-core material. The gas is gathered towards the massive protostar by the global collapse of the large clump as a whole.

\begin{figure}
\plotone{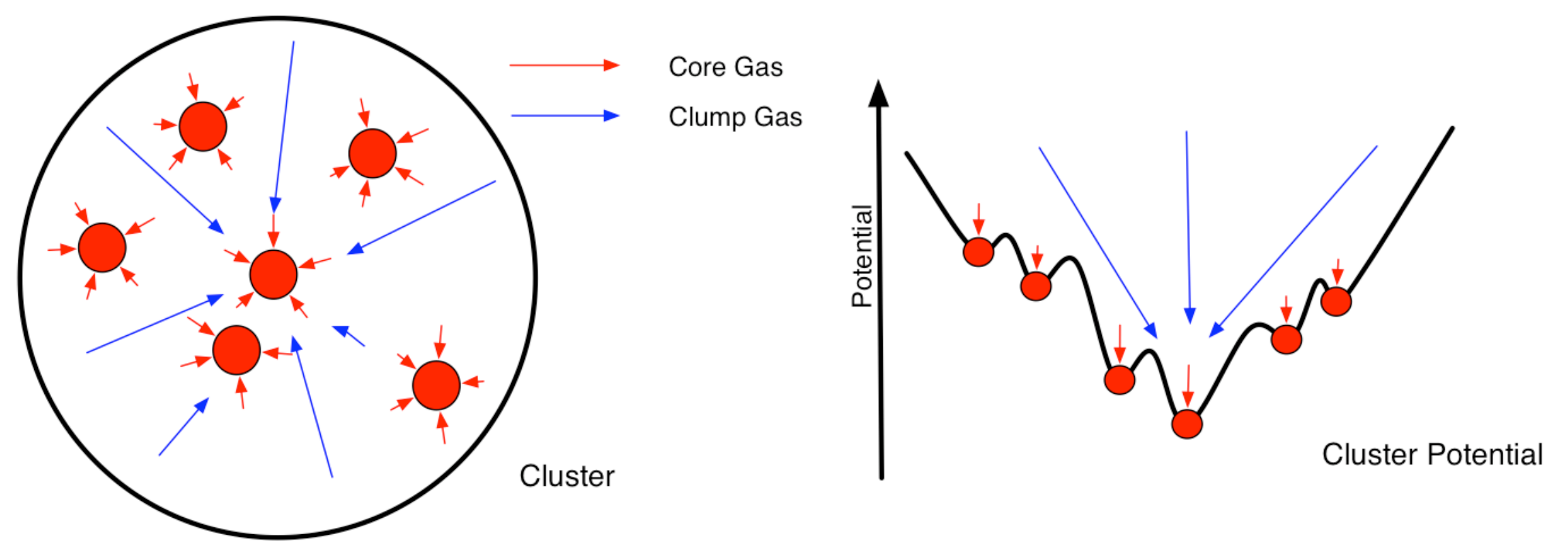}
\caption{A cartoon picture of how the gas becomes locked up into stars in a globally contracting cluster.
\label{fig:assembly}
}
\end{figure}

Figure \ref{fig:assembly} shows a cartoon picture of the how mass becomes partitioned into stars as the cluster evolves. On small scales, bound cores of gas collapse and convert their local mass into stellar systems. However, there is an additional larger-scale collapse as gas falls to the centre of the cluster potential. This allows cores located at the focal point of the large scale collapse to continue to accrete gas beyond the point at which their original core is consumed. For example, the central star of the cluster shown in Figure \ref{fig:fate} has a mass of $29.2$ M$_{\odot}$ at the end of the simulation, but its precursor core had a mass of only $0.67$ M$_{\odot}$. This is different from the original formulation of the competitive accretion theory \citep{Zinnecker82,Bonnell01} in which Bondi-Hoyle accretion was assumed, as the dynamics of this region are dominated by the gas potential rather than the stellar potential, and the central star is essentially stationary. Therefore the version of competitive accretion theory that assumes that tidal accretion dominates \citep{Bonnell01} is a better model for the growth of the massive central star. However, the mass function seen in these regions still follows a Salpeter like slope despite the fact that the tidally-dominated scenario predicts a somewhat shallower upper mass function. 

\section{Environmental Effects}

Given that most of the mass going into the massive stars actually comes from the diffuse cluster gas rather than dense cores, it is crucial that we capture this environment accurately in numerical simulations. In this section we shall look at three cases of previously ignored environmental physics and see if they change the upper end of the resulting stellar mass functions.

\subsection{Chemical evolution}
Generally in numerical studies of star formation the gas is either assumed to be isothermal, or is modeled by a polytropic equation of state \citep[e.g.][]{Larson05,Jappsen05}. However, in reality, the thermal evolution of gas within a molecular cloud is a time-dependent function of the cloud chemistry. Recently, detailed chemistry and cooling modules \citep{Glover07,Glover10} have been included in the Gadget 2 SPH code \citep{Springel05}. Processes within the thermodynamic model include photoelectric emission from dust grains, H$_2$ formation heating, excitation and photodissociation of H$_2$, cosmic ray ionisation, heating of dust by the interstellar radiation field, molecular and atomic line cooling, compressional and shock heating from dynamical processes, and time-dependent H$_2$ and CO formation and destruction.

\begin{figure}
\plottwo{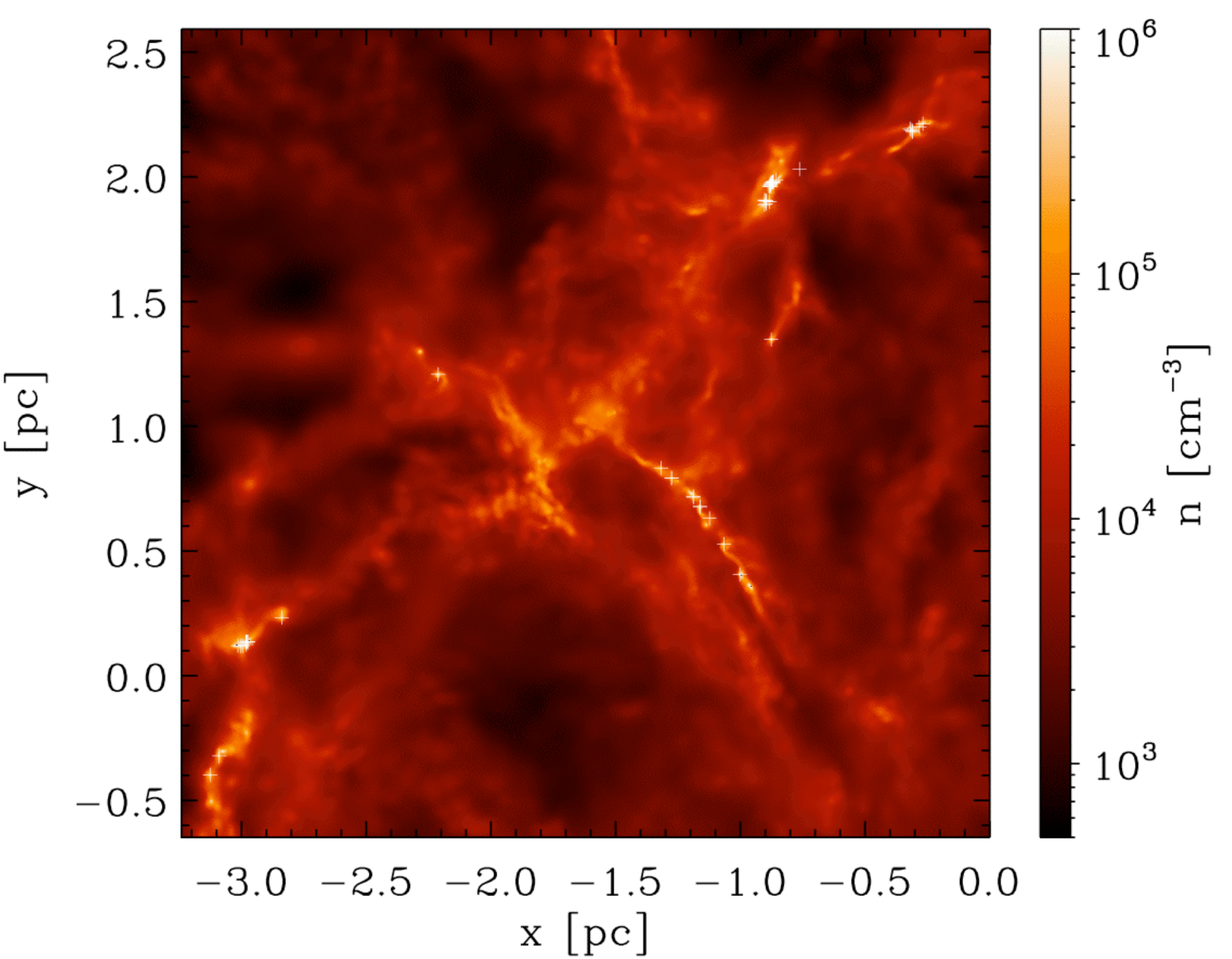}{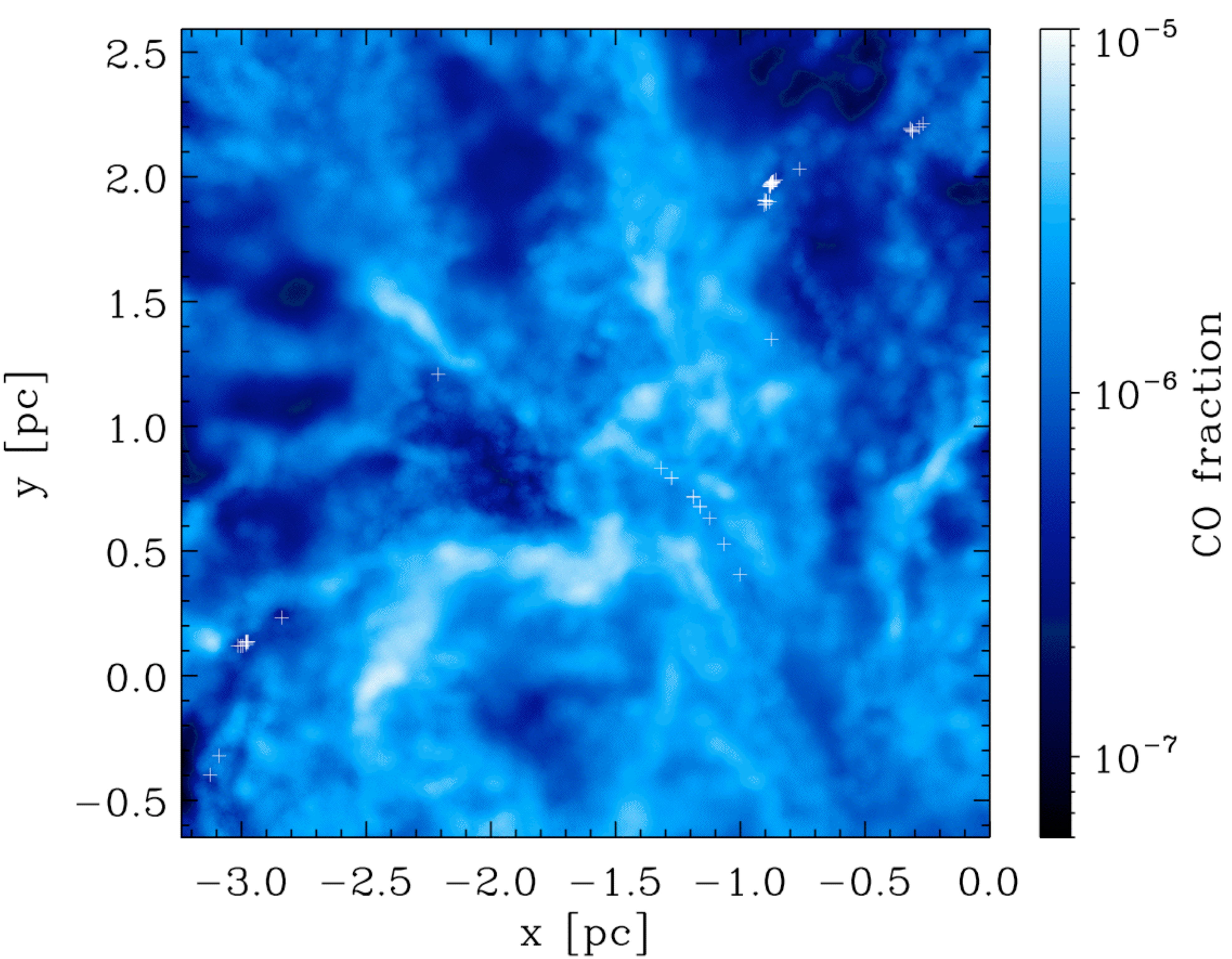}
\caption{\textit{Left} Maximum density along the line of sight at the onset of star formation. \textit{Right} the mean CO fraction of the gas. The maximum density and CO fraction are poorly correlated.
\label{fig:CO}
}
\end{figure}

This full chemical treatment has now been used to model the evolution of a $10^4$ M$_{\odot}$ turbulent molecular cloud irradiated by the standard interstellar radiation field (Clark et.~al., in preparation). Figure \ref{fig:CO} shows the maximum density along the line of sight and the CO fraction of the central region of the cloud at the onset of star formation. The two distributions are poorly correlated, as in cold dense regions the CO has been depleted onto dust grains. Another difference with previous models is that average temperatures of up to $50$ K are now seen in the more diffuse gas surrounding the star formation regions. This is due to photoelectric emission from dust grains and compressional heating of the gas. Nonetheless, despite these differences from previous studies, the actual star formation is broadly similar. This work is still at a provisional stage, but at present the upper end of the stellar mass function produced in the simulation is consistent with a Salpeter slope. The major advantage of this work is that it is now possible to directly compare between the simulations and real data using the actual observational quantities (Shetty et.~al., in prep).

\subsection{Gas metallicity}
A clear example of a situation in which the upper end of the mass function is thought to differ from the standard Salpeter slope is that of primordial star formation. The first stars are thought to be very massive \citep[see e.g.][]{BrommLarson04,Glover05,OShea07}, owing to the lack of effective coolants other than H$_{2}$ and HD in zero metalicity gas, which is expected to inhibit fragmentation. Moreover, it has been assumed that the first stars are single, as the envelopes in which they form show no sign of fragmentation up to the point at which the first star forms \citep[e.g.][]{Abel02,Yoshida06}. This would lead to a mass function resembling a delta function. However in more recent works some evidence of binarity has been seen \citep[][Clark et. al. in prep]{Turk09,Stacy10}, although this would still not produce a full mass function.

\begin{figure}
\plotone{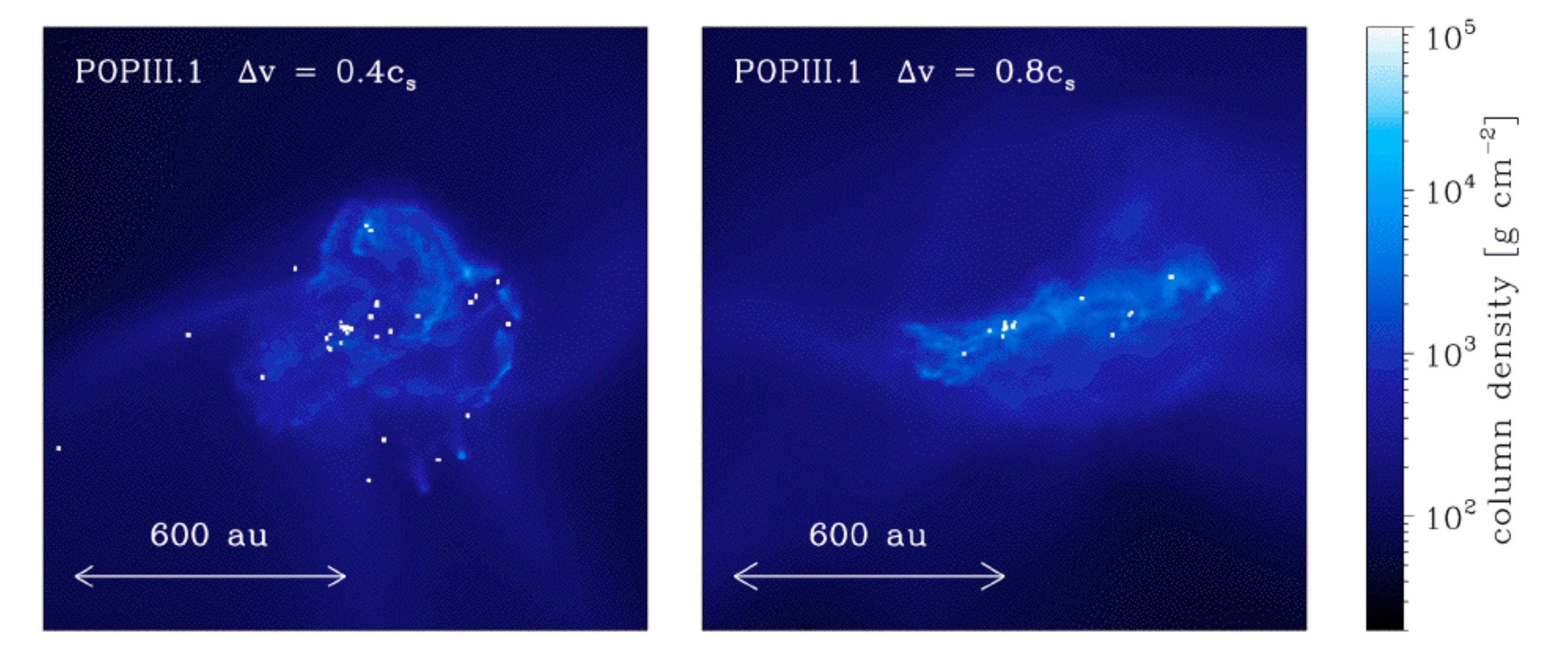}
\caption{Column density images showing the state of Pop.\ III clouds after they have converted 10 percent of their mass (100 M$_{\odot}$) into sinks. The sink particles are denoted in the images by the white dots, and in each case we centre the image on the first sink particle to form in the simulation. The level of turbulence is shown on each plot in terms of the initial sound speed c$_s$.
\label{fig:popIII}
}
\end{figure}

However, a recent study by \citet{Clark10} has shown that in the presence of mild subsonic turbulence, even primordial gas is highly susceptible to fragmentation. Figure \ref{fig:popIII} shows the central region of a collapsing Bonnor-Ebert sphere of primordial gas with subsonic turbulence. The sink particles formed in the haloes have masses in the range $0.1$ M$_{\odot}$ to $40$ M$_{\odot}$ and the fragmentation is dependent on the local properties of the turbulent velocity field, and hence is predicted to vary between star-forming mini-haloes. In conclusion, it is still to early to make predictions about the final shape of any primordial stellar mass function, but it may resemble the solar metallicity case more than previously expected.

\subsection{Unbound regions}
\begin{figure}
\begin{center}
\includegraphics[width=2in]{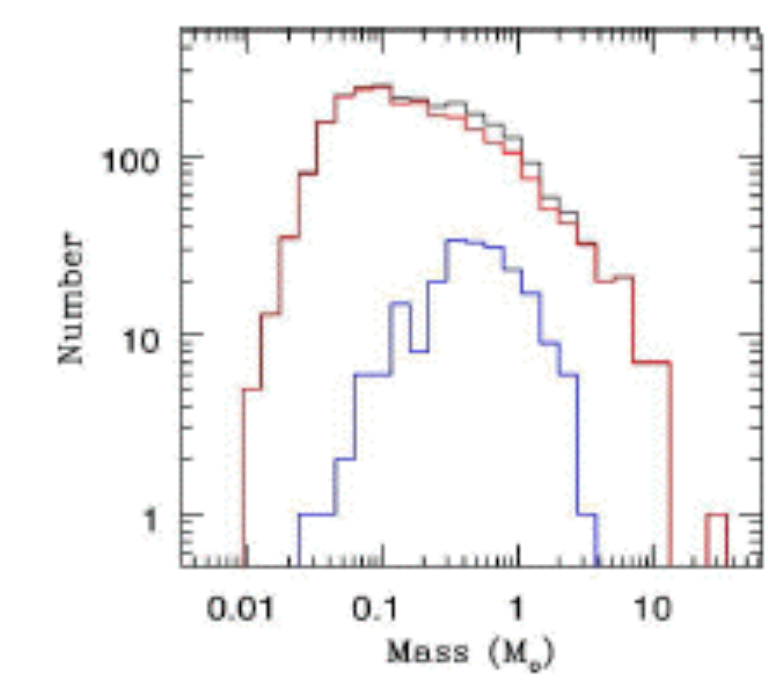}
\caption{The sink mass functions at the end of the simulation of the total (top, black line) sink population and for those
sinks formed in \textit{red }high-density regions (over 100 stars pc$^{-3}$) and \textit{blue} low density regions. The low density regions are unbound and do not form either high or low mass stars and thus produce an unusual IMF.}
\label{fig:unbound}
\end{center}
\end{figure}

In Section 1 we found that the process of global collapse was crucial to forming massive stars in dense stellar clusters. This raises the question; what happens when regions of the cloud are unbound and not undergoing collapse? \citet{Clark05a} showed that star formation still proceeds in globally unbound regions, but that the star forming efficiency is lower. Further to this, \citet{Clark08} and Bonnell et. al. (2010 \textit{submitted}) examine how the mass function differs in regions which have a low stellar density and are not globally bound in a simulated GMC. Figure \ref{fig:unbound} shows the sink mass function in regions with low and high stellar densities. The mass function in the unbound regions differs from the global mass function. Firstly, there are no massive stars as there is no global collapse (although since fewer stars form in this region, this result may also be expected as a consequence of random sampling of an invariant IMF). Secondly, the mass function has fewer low mass stars, since due to the lower stellar density there are fewer dynamical interactions and less tidal shear from the cluster environment during the protostars' evolution \citep{Bonnell08}. However, these effects will be quickly obscured due to stars from the nearby bound clusters being ejected into the unbound regions.

\section{Conclusions}
In summary, we conclude that:
\begin{enumerate}
\item Massive stars are formed from the global collapse of a larger region, rather than from a distinct pre-stellar core.
\item A treatment of the chemical evolution of molecular clouds produces qualitatively similar star formation behaviour to previous work, but facilitates a better comparison with observational data.
\item Even in primordial gas, fragmentation can produce a mass function.
\item The expected mass function of unbound regions in a molecular cloud is deficient in low and high mass stars compared to the standard IMF.
\end{enumerate}

\acknowledgements
RJS thanks the organisers of the UP2010 for an interesting and stimulating conference. We acknowledge financial support from the {\em Landesstiftung Baden-W{\"u}rttemberg} via their program International Collaboration II (grant P-LS-SPII/18) and from the German {\em Bundesministerium f\"{u}r Bildung und Forschung} via the ASTRONET project STAR FORMAT (grant 05A09VHA). The authors furthermore gives thanks for subsidies from the {\em Deutsche Forschungsgemeinschaft} (DFG) under grants no.\ KL 1358/1, KL 1358/4, KL 1359/5, KL 1358/10, and KL 1358/11, as well as from a Frontier grant of Heidelberg University sponsored by the German Excellence Initiative. This work was assisted by the European Commission FP6 Marie Curie
RTN CONSTELLATION (MRTN-CT-2006-035890) and was supported in part by the U.S. Department of Energy contract no. DE-AC-02-76SF00515. R.S.K. also thanks the KIPAC at Stanford University and the Department of Astronomy and Astrophysics at the University of California at Santa Cruz for their warm hospitality during a sabbatical stay in spring 2010.

\bibliography{Smith_R.bib}

\end{document}